\documentclass[aip,jmp,preprint,numerical]{revtex4-1}

\usepackage{amsmath}
\usepackage{amssymb}
\usepackage{amsfonts}
\usepackage{graphicx}
\usepackage{dcolumn}
\usepackage{bm}
\usepackage{hyperref}
\hypersetup{colorlinks=false}

\usepackage[dvips]{color}

\begin{document}

\title[Quantum Newtonian cosmology and the biconfluent Heun functions]{Quantum Newtonian cosmology and the biconfluent Heun functions}

\date{\today}

\author{H. S. Vieira}
\email{horacio.santana.vieira@hotmail.com}
\affiliation{Departamento de F\'{i}sica, Universidade Federal da Para\'{i}ba, Caixa Postal 5008, CEP 58051-970, Jo\~{a}o Pessoa, PB, Brazil}
\affiliation{Centro de Ci\^{e}ncias, Tecnologia e Sa\'{u}de, Universidade Estadual da Para\'{i}ba, CEP 58233-000, Araruna, PB, Brazil}
\author{V. B. Bezerra}
\email{valdir@fisica.ufpb.br}
\affiliation{Departamento de F\'{i}sica, Universidade Federal da Para\'{i}ba, Caixa Postal 5008, CEP 58051-970, Jo\~{a}o Pessoa, PB, Brazil}

\begin{abstract}
We obtain the exact solution of the Schr\"{o}dinger equation for a particle (galaxy) moving in a Newtonian universe with a cosmological constant, which is given in terms of the biconfluent Heun functions. The first six Heun polynomials of the biconfluent function are written explicitly. The energy spectrum which resembles the one corresponding to the isotropic harmonic oscillator is also obtained. The wave functions as well as the energy levels codify the role played by the cosmological constant.
\end{abstract}




\maketitle


%
%
\section{Introduction}
The equations of the Einsteinian cosmology are derived from General Relativity, with the use of the Cosmological Principle. This formulation is geometric, so that the descriptions of the cosmological effects are associated to the geometry of the spacetime, which is quantified by its curvature \cite{Schutz:2009}. In this case, differential geometry plays a fundamental role together with a wide range of some analytical tools which arise from geometry. In this metric approach, the models of the universe are then classified in accordance with the features of the curvature tensor.

In Newtonian cosmology, the equation that describes the evolution of the universe is obtained from the equation of motion and the energy equation for particles (galaxies) subject to gravitational forces and considering the Cosmological Principle as well \cite{Bondi:2010}. This suggests that we can use the formalism developed by Lagrange and Hamilton, writing down the Lagrangian and Hamiltonian of the system in order to obtain the equation and integral of motion.

The Newtonian cosmology, a model of the universe based on Newtonian mechanics, was well established in the third decade of last century \cite{QJMath.5.64,QJMath.5.73}. In this context, the cosmological equations which describe the time evolution of the universe obtained by considering the Cosmological Principle, assuming that the pressure is zero and using Newtonian dynamics and gravitation. The results obtained in this case show that the universe could not be static, obeying an equation which algebraically is similar to Friedmann's equation of General Relativity, but obviously has different interpretation. This model based on Newtonian treatment provides a good description of the expanding universe in which pressure is zero \cite{AmJPhys.33.105}. When pressure is taken into account it is necessary to modify the basic equations of the pressureless case \cite{ProcRSocLondA.206.562} and adopt some assumptions \cite{ProcRSocLondA.149.384} in order to garantee the analogy between the Newtonian and Einsteinian approaches.

It is worth call attention to the fact that to obtain the equivalence between Newtonian and Einsteinian cosmologies it is necessary to combine the equations of classical mechanics and the Cosmological Principle which nowdays is strongly supported by the smooth Cosmic Microwave Background Radiation (CMB) whose spectrum is isotropic and uniform to \cite{PhysRevLett.109.051303,AstronAstrophys.571.A15} within $(\Delta T/T) \leq 10^{-5}$. Thus, the fact that this anisotropy agrees very well with that of the initially homogeneous and isotropic universe indicates that the Cosmological Principle should be valid at least over length scales much bigger than hundreds of Megaparsecs.

Quantum Newtonian cosmology is the theoretical framework within which it is possible to find the wave function which predicts the behavior of the Newtonian universe. This wave function for Newtonian cosmology is obtained in non-relativistic quantum mechanics, and thus, all we need is to solve the Schr\"{o}dinger equation for the system under consideration \cite{AIPConfProc.743.286,arXiv:0504072,ProcRSoc.A.463.503,IntJTheorPhys.47.455,ISRNMathPhys.2013.509316}.

A Heun equation is a second-order linear differential equation with four regular sin\-gu\-lar\-i\-ties \cite{MathAnn.33.161}. The confluent form of the Heun equation is obtained when two singular points coalesce and the point at infinity becomes irregular. In this case, we have the biconfluent Heun equation \cite{JPhysAMathGen.19.3527,AnnPhys.347.130}. The Heun functions have a wide range of applications in physics, from quantum mechanics to cosmology \cite{arXiv:1101.0471v5} and the number of articles using their solutions have been growing, even in a scenario in which the theory concerning these functions is not yet completed.

This paper is organized as follows. In section 2 we obtain the Hamiltonian operator for the Newtonian universe. In section 3, we solve the Schr\"{o}dinger equation in this background. We present the biconfluent Heun equation, and then we obtain the Heun polynomials in section 4. In section 5 we obtain the energy spectrum for a particle (galaxy) moving in a Newtonian universe. In section 6, we write the general and exact expression of the wave function. Finally, in section 7, we present our conclusions.
%
%
\section{Hamiltonian operator}
In our previous paper \cite{RevBrasEnsFis.36.3310}, using some assumptions about the structure of the Newtonian universe, we obtained the Lagrangian, $L$, for the motion of a particle (galaxy) of mass $m$, which is written as
\begin{equation}
L(R,\dot{R})=\frac{1}{2}m\dot{R}^{2}+\frac{GMm}{R}+\frac{1}{6}\Lambda mR^{2}\ ,
\label{eq:Lagrangeana}
\end{equation}
where $\Lambda$ is the cosmological constant, and the last term in Eq.~(\ref{eq:Lagrangeana}) corresponds to a kind of cosmological potential energy. Using this Lagrangian, the equation for the scale parameter, $R(t)$, can be written as
\begin{equation}
\ddot{R}=-\frac{4}{3}\pi G\rho R+\frac{1}{3}\Lambda R\ ,
\label{eq:cosmologica}
\end{equation}
where $M=4\pi R^{3}\rho/3$ is the mass of the sphere (total mass of the Newtonian universe). It is worth call attention to the fact that we are considering the case $p=0$, that is, a universe in which the pressure is zero.

The Lagrangian given by Eq.~(\ref{eq:Lagrangeana}) does not depend explicitly on time, because the system is under the action of a uniform force field, so that the constant of motion, $H_{cl}$, called classical Hamiltonian of the system, can be defined as
\begin{equation}
H_{cl}=\left(\frac{\partial L}{\partial\dot{R}}\right)\dot{R}-L\ .
\label{eq:Hamiltoniano}
\end{equation}
From Eq.~(\ref{eq:Hamiltoniano}), we can obtain an equation for the scale factor, $R(t)$, which is analogous to the Friedmann equation, whose form is
\begin{equation}
\dot{R}^{2}=\frac{C}{R}+\frac{1}{3}\Lambda R^{2}-k\ ,
\label{eq:dif_cosmologica}
\end{equation}
where $C=8 \pi G \rho R^{3}/3$ and $k=-2E/m$ are constants, and $E=H_{cl}$ is the total energy. Thus, from Eq.~(\ref{eq:Hamiltoniano}) we can write the classical Hamiltonian as
\begin{equation}
H_{cl}(R,\dot{R})=\frac{1}{2}m\dot{R}^{2}-\frac{GMm}{R}-\frac{1}{6}\Lambda mR^{2}\ .
\label{eq:Hamiltonian}
\end{equation}

Now, we can construct a phase space as $(R,P_{R})$, where $R$ is the comoving coordinate with
\begin{equation}
P_{R}=m\dot{R}
\label{eq:Momentum}
\end{equation}
being the linear momentum. Thus, substituting Eq.~(\ref{eq:Momentum}) into Eq.~(\ref{eq:Hamiltonian}), the classical Hamiltonian takes the form
\begin{equation}
H_{cl}(R,P_{R})=\frac{P^{2}_{R}}{2m}-\frac{GMm}{R}-\frac{1}{6}\Lambda mR^{2}\ .
\label{eq:Hamiltonian_op}
\end{equation}
Therefore, performing the canonical quantization
\begin{equation}
P_{R}\ \rightarrow\ P_{R_{op}} = -i\hbar\frac{d}{dR}\ ,
\label{eq:Canonical_quant}
\end{equation}
we obtain the Hamiltonian operator for a particle (galaxy) moving in the Newtonian universe, which is given by
\begin{equation}
H=-\frac{\hbar^{2}}{2m}\frac{d^{2}}{dR^{2}}-\frac{GMm}{R}-\frac{1}{6}\Lambda mR^{2}\ .
\label{eq:Hamiltonian_op-canonical}
\end{equation}
%
%
\section{Schr\"{o}dinger equation}
In this section we will solve the Schr\"{o}dinger equation for a particle (galaxy) in the Newtonian universe, given by
\begin{equation}
H\psi(R)=E\psi(R)\ ,
\label{eq:Schrodinger}
\end{equation}
where it is understood that the eigenfunction $\psi(R)$ corresponds to the eigenvalue $E$, with $\Psi(R,t)=\psi(R)\mbox{e}^{-iEt/\hbar}$ being the general wave function of the time-dependent Schr\"{o}dinger equation.

Substituting Eq.~(\ref{eq:Hamiltonian_op-canonical}) into Eq.~(\ref{eq:Schrodinger}), we obtain
\begin{equation}
-\frac{\hbar^{2}}{2m}\frac{d^{2}\psi(R)}{dR^{2}}+\left(-\frac{GMm}{R}-\frac{1}{6}\Lambda mR^{2}\right)\psi(R)=E\psi(R)\ .
\label{eq:mov_1}
\end{equation}
Note that all eigenfunctions correspond to bound states of positive energy. It is more suitable to rewrite Eq.~(\ref{eq:mov_1}) in terms of dimensionless quantities. To do this, we first introduce the dimensionless parameter
\begin{equation}
\gamma=\frac{2E}{\hbar\Omega}\ ,
\label{eq:gamma_mov_1}
\end{equation}
where
\begin{equation}
\Omega=\left(-\frac{\Lambda}{3}\right)^{1/2}\ .
\label{eq:Omega_mov_1}
\end{equation}
Let us also use the dimensionless variable
\begin{equation}
x=\tau R\ ,
\label{eq:x}
\end{equation}
where the coefficient $\tau$ is given by
\begin{equation}
\tau=\left(\frac{m\Omega}{\hbar}\right)^{1/2}\ .
\label{eq:tau_mov_1}
\end{equation}
Then, Eq.~(\ref{eq:mov_1}) turns into
\begin{equation}
\frac{d^{2}\psi(x)}{dx^{2}}+\left(\gamma-x^{2}-\frac{1}{2}\delta\frac{1}{x}\right)\psi(x)=0\ ,
\label{eq:mov_2}
\end{equation}
where the parameter $\delta$ is given by
\begin{equation}
\delta=-\frac{4GMm^{2}}{\tau\hbar^{2}}\ .
\label{eq:delta_mov_2}
\end{equation}

Now, let us analyse the asymptotic behaviour of $\psi$ when $|x| \rightarrow \infty$. For any finite value of the total energy $E$, the quantity $\gamma$ and the term proportional to $x^{-1}$ becomes negligible with respect to $x^{2}$, in the limit $|x| \rightarrow \infty$, so that in this limit, Eq.~(\ref{eq:mov_2}) reduces to
\begin{equation}
\left(\frac{d^{2}}{dx^{2}}-x^{2}\right)\psi(x)=0\ .
\label{eq:mov_3}
\end{equation}
The functions which satisfy this equation for large enough $|x|$ are
\begin{equation}
\psi(x)=x^{r}\mbox{e}^{\pm x^{2}/2}\ .
\label{eq:approx_sol_mov_3}
\end{equation}
The wave function $\psi$ must be bounded everywhere, including $x=\pm\infty$. Therefore, the physically acceptable solutions must contain only the minus sign in the exponent. The analysis of the asymptotic behaviour of the system under consideration suggests that we have to look for solutions of Eq.~(\ref{eq:mov_2}) valid for all $x$ and having the form
\begin{equation}
\psi(x)=x\ \mbox{e}^{-x^{2}/2}P(x)\ ,
\label{eq:form_sol_mov_3}
\end{equation}
where $P(x)$ is to be determined appropriatelly. Substituting Eq.~(\ref{eq:form_sol_mov_3}) into Eq.~(\ref{eq:mov_2}), we conclude that $P(x)$ must satisfy the equation
\begin{equation}
\frac{d^{2}P}{dx^{2}}+\left(\frac{2}{x}-2x\right)\frac{dP}{dx}+\left[(\gamma-3)-\frac{1}{2}\delta\frac{1}{x}\right]P=0\ ,
\label{eq:mov_4}
\end{equation}
whose analytical solutions will be obtained in the next Section.
%
%
\section{Biconfluent Heun equation}
Equation (\ref{eq:mov_4}) is a particular case of the biconfluent Heun equation \cite{Ronveaux:1995}, which in the canonical form is given by
\begin{eqnarray}
&& \frac{d^{2}y}{dx^{2}}+\left(\frac{1+\alpha}{x}-\beta-2x\right)\frac{dy}{dx}\nonumber\\
&& + \left\{(\gamma-\alpha-2)-\frac{1}{2}[\delta+(1+\alpha)\beta]\frac{1}{x}\right\}y=0\ ,
\label{eq:Biconfluent_Heun_Canonical}
\end{eqnarray}
where $y(x)=\mbox{HeunB}(\alpha,\beta,\gamma,\delta;x)$ are the biconfluent Heun functions. 

Let us assume that the solutions of Eq.~(\ref{eq:Biconfluent_Heun_Canonical}) can be written as
\begin{equation}
y(x)=\sum_{s=0}^{\infty}c_{s}x^{s}\ ,
\label{eq:y(x)_Biconfluent_Heun}
\end{equation}
and thus, we have
\begin{equation}
\frac{d}{dx}y(x)=\sum_{s=1}^{\infty}sc_{s}x^{s-1}\ ,
\label{eq:y'(x)_Biconfluent_Heun}
\end{equation}
\begin{equation}
\frac{d^{2}}{dx^{2}}y(x)=\sum_{s=2}^{\infty}s(s-1)c_{s}x^{s-2}\ .
\label{eq:y''(x)_Biconfluent_Heun}
\end{equation}
Substituting Eqs.~(\ref{eq:y(x)_Biconfluent_Heun}), (\ref{eq:y'(x)_Biconfluent_Heun}) and (\ref{eq:y''(x)_Biconfluent_Heun}) into Eq.~(\ref{eq:Biconfluent_Heun_Canonical}), we arrive at
\begin{eqnarray}
&& \sum_{s=2}^{\infty}s(s-1)c_{s}x^{s-2}+\left(\frac{1+\alpha}{x}-\beta-2x\right)\sum_{s=1}^{\infty}sc_{s}x^{s-1}\nonumber\\
&& + \left\{(\gamma-\alpha-2)-\frac{1}{2}[\delta+(1+\alpha)\beta]\frac{1}{x}\right\}\sum_{s=0}^{\infty}c_{s}x^{s}=0\ ,
\label{eq:Biconfluent_Heun_Canonical_expansion_1}
\end{eqnarray}
from which follows
\begin{eqnarray}
&& \sum_{s=2}^{\infty}s(s-1)c_{s}x^{s-2}+\sum_{s=1}^{\infty}(1+\alpha)sc_{s}x^{s-2}-\sum_{s=1}^{\infty}\beta sc_{s}x^{s-1}\nonumber\\
&& - \sum_{s=1}^{\infty}2sc_{s}x^{s}+\sum_{s=0}^{\infty}(\gamma-\alpha-2)c_{s}x^{s}-\sum_{s=0}^{\infty}\frac{1}{2}[\delta+(1+\alpha)\beta]c_{s}x^{s-1}=0\ ,\nonumber\\
\label{eq:Biconfluent_Heun_Canonical_expansion_2}
\end{eqnarray}
or, equivalently
\begin{eqnarray}
&& \sum_{S=0}^{\infty}(S+2)(S+1)c_{S+2}x^{S}+\sum_{S=-1}^{\infty}(1+\alpha)(S+2)c_{S+2}x^{S}\nonumber\\
&& - \sum_{S=0}^{\infty}\beta (S+1)c_{S+1}x^{S}-\sum_{S=1}^{\infty}2Sc_{S}x^{S}+\sum_{S=0}^{\infty}(\gamma-\alpha-2)c_{S}x^{S}\nonumber\\
&& - \sum_{S=-1}^{\infty}\frac{1}{2}[\delta+(1+\alpha)\beta]c_{S+1}x^{S}=0\ ,
\label{eq:Biconfluent_Heun_Canonical_expansion_3}
\end{eqnarray}
where $S=s-2$. Collecting all terms of same order in $x$, we get
\begin{eqnarray}
&& \left\{(1+\alpha)c_{1}-\frac{1}{2}[\delta+(1+\alpha)\beta]c_{0}\right\}x^{-1}\nonumber\\
&& + \{ 2(2+\alpha)c_{2}-\left\{\frac{1}{2}[\delta+(1+\alpha)\beta]+\beta\right\}c_{1}+(\gamma-\alpha-2)c_{0} \} \nonumber\\
&& + \sum_{S=1}^{\infty} \{ (S+2)(S+2+\alpha)c_{S+2}-\left\{\frac{1}{2}[\delta+(1+\alpha)\beta]+(S+1)\beta\right\}c_{S+1}\nonumber\\
&& + (\gamma-\alpha-2-2S)c_{S} \} x^{S}=0\ .
\label{eq:Biconfluent_Heun_Canonical_expansion_4}
\end{eqnarray}

Thus, Eq.~(\ref{eq:Biconfluent_Heun_Canonical_expansion_4}) gives the following recursion relations for the expansion coefficients
\begin{eqnarray}
	(1+\alpha)c_{1}=\frac{1}{2}[\delta+(1+\alpha)\beta]c_{0}\ ,\nonumber\\
	2(2+\alpha)c_{2}=\left\{\frac{1}{2}[\delta+(1+\alpha)\beta]+\beta\right\}c_{1}-(\gamma-\alpha-2)c_{0}\ ,\nonumber\\
	(S+2)(S+2+\alpha)c_{S+2}=\left\{\frac{1}{2}[\delta+(1+\alpha)\beta]+(S+1)\beta\right\}c_{S+1}\nonumber\\
-(\gamma-\alpha-2-2S)c_{S}, \quad S \geq 0\ .
\label{eq:recursion_Biconfluent_Heun_expansion_4}
\end{eqnarray}
Considering an appropriate choice of $c_{0}$, namely $c_{0}=1$, relations given by Eq.~(\ref{eq:recursion_Biconfluent_Heun_expansion_4}) turn into
\begin{eqnarray}
	c_{0}=1\ ,\nonumber\\
	(1+\alpha)c_{1}=\frac{1}{2}[\delta+(1+\alpha)\beta]\ ,\nonumber\\
	2(1+\alpha)(2+\alpha)c_{2}=\left\{\frac{1}{2}[\delta+(1+\alpha)\beta]+\beta\right\}\frac{1}{2}[\delta+(1+\alpha)\beta]\nonumber\\
-(1+\alpha)(\gamma-\alpha-2)\ ,\nonumber\\
	6(1+\alpha)(2+\alpha)(3+\alpha)c_{3}=\left\{\frac{1}{2}[\delta+(1+\alpha)\beta]+2\beta\right\}\nonumber\\
\times \{ \left\{\frac{1}{2}[\delta+(1+\alpha)\beta]+\beta\right\}\frac{1}{2}[\delta+(1+\alpha)\beta]-(1+\alpha)(\gamma-\alpha-2) \} \nonumber\\
-2(2+\alpha)(\gamma-\alpha-4)\frac{1}{2}[\delta+(1+\alpha)\beta]\ .
\label{eq:recursion_c_Biconfluent_Heun_expansion_4}
\end{eqnarray}
Thus, taking into account a convenient change in notation, we have
\begin{eqnarray}
	A_{0}=c_{0}=1\ ,\nonumber\\
	A_{1}=(1+\alpha)c_{1}=\frac{1}{2}[\delta+(1+\alpha)\beta]\ ,\nonumber\\
	A_{2}=2(1+\alpha)(2+\alpha)c_{2}\ ,\nonumber\\
	A_{3}=6(1+\alpha)(2+\alpha)(3+\alpha)c_{3}\ ,\nonumber\\
	\vdots\nonumber\\
	A_{s}=s!(1+\alpha)_{s}c_{s}, \quad s \geq 0\ ,
\label{eq:recursion_A_Biconfluent_Heun_expansion_4}
\end{eqnarray}
where
\begin{eqnarray}
(1+\alpha)_{s} & = & \frac{\Gamma(s+1+\alpha)}{\Gamma(1+\alpha)}\nonumber\\
& = & \left\{
\begin{array}{ll}
	(1+\alpha)(2+\alpha)\ldots(s+\alpha) & \quad s=1,2,3,\ldots\\
	1 & \quad s=0\ .
\end{array}
\right.
\label{eq:alpha_s_Biconfluent_Heun_expansion}
\end{eqnarray}

Now, if we assume that $\alpha$ is not a negative integer, the biconfluent Heun functions can be written as \cite{AnnSocSciBruxelles.92.151}
\begin{equation}
\mbox{HeunB}(\alpha,\beta,\gamma,\delta;x)=\sum_{s \geq 0}\frac{A_{s}}{(1+\alpha)_{s}}\frac{x^{s}}{s!}\ ,
\label{eq:Biconfluent_Heun_expansion}
\end{equation}
where
\begin{eqnarray}
&& A_{s+2}=\left\{(s+1)\beta+\frac{1}{2}[\delta+(1+\alpha)\beta]\right\}A_{s+1}\nonumber\\
&& -(s+1)(s+1+\alpha)(\gamma-\alpha-2-2s)A_{s}, \quad s \geq 0\ .
\label{eq:recursion_Biconfluent_Heun_expansion}
\end{eqnarray}
%
%
\subsection{Heun polynomials}
From the recursion relation given by Eq.~(\ref{eq:recursion_Biconfluent_Heun_expansion}), the function $\mbox{HeunB}(\alpha,\beta,\gamma,\delta;x)$ becomes a polynomial of degree $n$ if and only if the two following conditions are fulfilled \cite{BullSocRSciLiege.40.13}:
\begin{equation}
\begin{array}{rl}
	\mbox{(i)} & \gamma-\alpha-2=2n, \quad n=0,1,2,\ldots\\
	\mbox{(ii)} & A_{n+1}=0
\end{array}
\label{eq:condiction_poly_Biconfluent_Heun}
\end{equation}
where $A_{n+1}$ is a polynomial of degree $n+1$ in $\delta$. Note that there are at most $n+1$ suitable values of $\delta$, labeled
\begin{equation}
\delta_{\mu}^{n}, \quad 0 \leq \mu \leq n\ .
\label{eq:delta_n_mu_Biconfluent_Heun}
\end{equation}

The polynomial $A_{n+1}$, which has $n+1$ real roots when $1+\alpha > 0$ and $\beta \in \mathbb{R}$, is the determinant of dimension $n+1$ given by
\begin{equation}
\left|
\begin{array}{ccccccc}
	\delta' & 1 & 0 & 0 & \ldots & \ldots & 0 \\
	2(1+\alpha)n & \delta'-\beta & 1 & 0 & \ldots & \ldots & 0 \\
	0 & 4(2+\alpha)(n-1) & \delta'-2\beta & 1 & 0 & \ldots & 0 \\
	0 & 0 & \gamma_{2} & \delta'-3\beta & 1 & \ldots & 0 \\
	\vdots & \vdots & 0 & \ddots & \ddots & \ddots & \ldots \\
	\vdots & \vdots & \vdots & \vdots & \gamma_{s-1} & \delta'_{s-1} & 1 \\
	0 & 0 & 0 & 0 & 0 & \gamma_{s} & \delta'_{s}
\end{array}
\right|\ ,
\label{eq:determinant_Biconfluent_Heun}
\end{equation}
where
\begin{equation}
\delta' \equiv -\frac{1}{2}[\delta+(1+\alpha)\beta]\ ,
\label{eq:delta'}
\end{equation}
\begin{equation}
\delta'_{s}=\delta'-(s+1)\beta\ ,
\label{eq:}
\end{equation}
\begin{equation}
\gamma_{s}=2(s+1)(s+1+\alpha)(n-s)\ .
\label{eq:gamma_s}
\end{equation}

Now, let us return to the wave functions $\psi(x)$. Using Eq.~(\ref{eq:form_sol_mov_3}), we see that the physically acceptable solutions of Eq.~(\ref{eq:mov_4}) are given by
\begin{equation}
\psi_{n,\mu}(x)=x\ \mbox{e}^{-x^{2}/2}P_{n,\mu}(\alpha,\beta;x)\ ,
\label{eq:form_polyn_sol_mov_3}
\end{equation}
where the functions $P_{n,\mu}(\alpha,\beta;x)$ are polynomials of degree $n$ satisfying the biconfluent Heun equation given by Eq.~(\ref{eq:Biconfluent_Heun_Canonical}), with $\gamma=\alpha+2(n+1)$, $A_{n+1}=0$, and $0 \leq \mu \leq n$, namely:
\begin{equation}
\frac{d^{2}P_{n,\mu}}{dx^{2}}+\left(\frac{1+\alpha}{x}-\beta-2x\right)\frac{dP_{n,\mu}}{dx}+\left\{2n-\frac{1}{2}[\delta_{\mu}^{n}+(1+\alpha)\beta]\frac{1}{x}\right\}P_{n,\mu}=0\ .
\label{eq:Biconfluent_Heun_polynomials}
\end{equation}

The polinomials $P_{n,\mu}(\alpha,\beta;x)=\mbox{HeunB}(\alpha,\beta,\alpha+2(n+1),\delta_{\mu}^{n};x)$ are called Heun polynomials of the biconfluent case. It is clear from the foregoing discussion that they are uniquely defined, except for an arbitrary multiplicative constant. Its highest order terms are given by
\begin{eqnarray}
&& x^{n}+\frac{1}{(1+\alpha)_{1}}\delta'\frac{x^{n-1}}{1!}+\frac{1}{(1+\alpha)_{2}}[\delta'(\delta'-\beta)-2(1+\alpha)n]\frac{x^{n-2}}{2!}\nonumber\\
&& + \frac{1}{(1+\alpha)_{3}}[\delta'(\delta'-\beta)(\delta'-2\beta)-4\delta'(2+\alpha)(n-1)\nonumber\\
&& -2(\delta'-2\beta)(1+\alpha)n]\frac{x^{n-3}}{3!}+Q(x)\ ,
\label{eq:polynomial_Biconfluent_Heun}
\end{eqnarray}
where $Q(x)$ is a polynomial with degree $\leq n-4$.

In our case, the parameters $\alpha$ and $\beta$ take the values $1$ and $0$, respectively. Therefore, the explicit form of the Heun polynomials $P_{n,\mu}(1,0;x)$, for $n=0,1,2$, are as follows:
\begin{itemize}
	\item $n=0$, $\gamma=3$, $\delta_{0}^{0}=0$
\end{itemize}
\begin{equation}
P_{0,0}(1,0;x)=1\ ;
\label{eq:P0}
\end{equation}

\begin{itemize}
	\item $n=1$, $\gamma=5$, $(\delta_{\mu}^{1})^{2}-16=0\ \Rightarrow\ \delta_{0}^{1}=+4,\delta_{1}^{1}=-4$
\end{itemize}
\begin{equation}
P_{1,0}(1,0;x)=x-\frac{1}{2}\frac{\delta_{0}^{1}}{2}=x-1\ ,
\label{eq:P1,0}
\end{equation}
\begin{equation}
P_{1,1}(1,0;x)=x-\frac{1}{2}\frac{\delta_{1}^{1}}{2}=x+1\ ;
\label{eq:P1,1}
\end{equation}

\begin{itemize}
	\item $n=2$, $\gamma=7$, $(\delta_{\mu}^{2})^{3}-80\delta_{\mu}^{2}=0\ \Rightarrow\ \delta_{0}^{2}=+4\sqrt{5},\delta_{1}^{2}=0,\delta_{2}^{2}=-4\sqrt{5}$
\end{itemize}
\begin{equation}
P_{2,0}(1,0;x)=x^{2}-\frac{1}{2}\frac{\delta_{0}^{2}}{2}x+\frac{1}{6}\left[\frac{(\delta_{0}^{2})^{2}}{4}-8\right]\frac{1}{2}=x^{2}-\sqrt{5}x+1\ ,
\label{eq:P2,0}
\end{equation}
\begin{equation}
P_{2,1}(1,0;x)=x^{2}-\frac{1}{2}\frac{\delta_{1}^{2}}{2}x+\frac{1}{6}\left[\frac{(\delta_{1}^{2})^{2}}{4}-8\right]\frac{1}{2}=x^{2}-\frac{2}{3}\ ,
\label{eq:P2,1}
\end{equation}
\begin{equation}
P_{2,2}(1,0;x)=x^{2}-\frac{1}{2}\frac{\delta_{2}^{2}}{2}x+\frac{1}{6}\left[\frac{(\delta_{2}^{2})^{2}}{4}-8\right]\frac{1}{2}=x^{2}+\sqrt{5}x+1\ .
\label{eq:P2,2}
\end{equation}

It is worth noting that for all values of $n$, namely, $n=0,1,2,\ldots$, correspond a function $\mbox{HeunB}(\alpha,\beta,\gamma,\delta;x)$ which is a polynomial of degree $n$ in $x$.
%
%
\section{Energy levels}
In order to obtain the energy spectrum, let us use Eqs.~(\ref{eq:gamma_mov_1}) and (\ref{eq:condiction_poly_Biconfluent_Heun}), and take into account that $\alpha=1$. Thus, the energy spectrum for a particle (galaxy) in the Newtonian universe is given by
\begin{equation}
E_{n}=\left(n+\frac{3}{2}\right)\hbar\Omega=\left(n+\frac{3}{2}\right)\hbar\left(-\frac{\Lambda}{3}\right)^{\frac{1}{2}}, \quad n=0,1,2,\ldots\ .
\label{eq:Energy_levels_Schrodinger_general}
\end{equation}

\begin{figure}
		\includegraphics[scale=0.40]{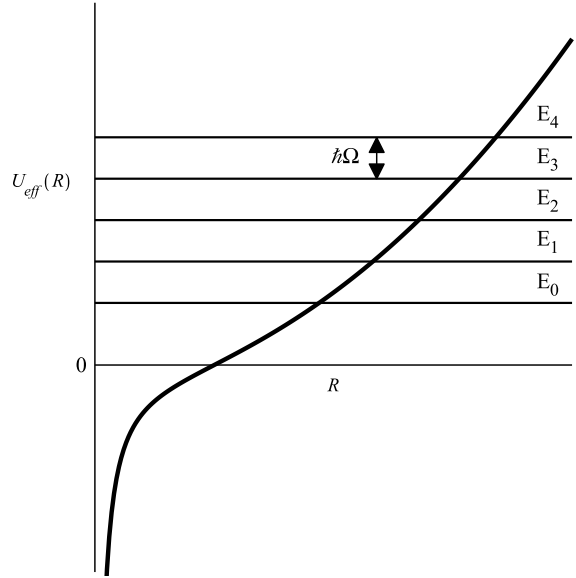}
	\caption{The effective potential energy of the Newtonian universe $U_{eff}(R)=-\frac{GMm}{R}+\frac{1}{6}|\Lambda| mR^{2}$ and the first five energy levels are shown.}
	\label{fig:Fig1}
\end{figure}

We see from Eq.~(\ref{eq:Energy_levels_Schrodinger_general}) that this quantum mechanical energy spectrum consist of an infinite sequence of discrete levels (see Fig.~\ref{fig:Fig1}), which are equally spaced and are similar to those obtained for the three-dimensional isotropic harmonic oscillator. Note that the eigenvalues given by Eq.~(\ref{eq:Energy_levels_Schrodinger_general}) are degenerate.

If we consider a scenario in which the cosmological constant is negative, that is, $\Lambda=-|\Lambda|$, we can rewritten the energy spectrum as
\begin{equation}
E_{n}=\left(n+\frac{3}{2}\right)\hbar\left(\frac{|\Lambda|}{3}\right)^{\frac{1}{2}}, \quad n=0,1,2,\ldots\ ,
\label{eq:Energy_levels_Schrodinger}
\end{equation}
where $(|\Lambda|/3)^{1/2}$ plays the role of the frequency if we compare Eq.~(\ref{eq:Energy_levels_Schrodinger}) with the corresponding to the energy spectrum of the three-dimensional isotropic harmonic oscillator.
%
%
\section{The wave function for Newtonian cosmology}
\begin{figure}
		\includegraphics[scale=0.45]{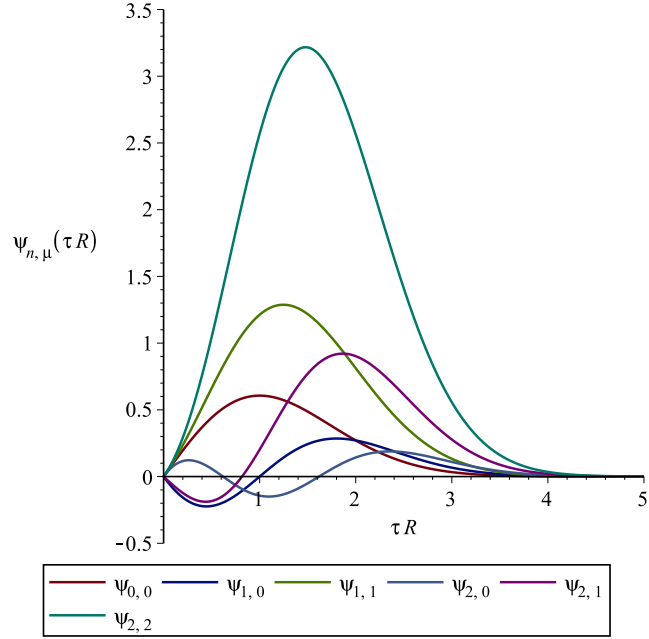}
	\caption{The first six wave functions of the Newtonian universe.}
	\label{fig:Fig2}
\end{figure}

Using Eq.~(\ref{eq:form_polyn_sol_mov_3}), we see that to each discrete value $E_{n}$, given by Eq.~(\ref{eq:Energy_levels_Schrodinger_general}), there corresponds $n+1$ physically acceptable eigenfunctions (see Fig.~\ref{fig:Fig2}), given by
\begin{eqnarray}
\psi_{n,\mu}(R) & = & N_{n}\ \tau R\ \mbox{e}^{-\tau^{2}R^{2}/2}\ P_{n,\mu}(\alpha,\beta;\tau R)\nonumber\\
& = & N_{n}\ \tau R\ \mbox{e}^{-\tau^{2}R^{2}/2}\ \mbox{HeunB}(\alpha,\beta,\alpha+2(n+1),\delta_{\mu}^{n};\tau R)\ ,
\label{eq:form_polyn_sol_mov_4}
\end{eqnarray}
where the parameters $\alpha$, $\beta$, $\gamma$, and $\delta$ are given by the following values and expressions:
\begin{equation}
\alpha=1\ ;
\label{eq:alpha_Schrodinger}
\end{equation}
\begin{equation}
\beta=0\ ;
\label{eq:beta_Schrodinger}
\end{equation}
\begin{equation}
\gamma=\frac{2E}{\hbar}\left(-\frac{3}{\Lambda}\right)^{\frac{1}{2}}\ ;
\label{eq:gamma_Schrodinger}
\end{equation}
\begin{equation}
\delta=-4GM\Biggl(\frac{m}{\hbar}\Biggr)^{3/2}\left(-\frac{3}{\Lambda}\right)^{1/4}\ .
\label{eq:delta_Schrodinger}
\end{equation}

The quantity $N_{n}$ in Eq.~(\ref{eq:form_polyn_sol_mov_4}) is a constant which (apart from a arbitrary phase factor) can be determined by requiring that the wave function given by Eq.~(\ref{eq:form_polyn_sol_mov_4}) obeys the following relation
\begin{equation}
\int_{0}^{\infty}|\psi_{n,\mu}(R)|^{2}dR=\frac{|N_{n}|^{2}}{\tau}\int_{0}^{\infty}x^{2}\mbox{e}^{-x^{2}}P_{n,\mu}^{2}(\alpha,\beta;x)dx=1\ .
\label{eq:normalization}
\end{equation}

In order to evaluate this integral, we consider the power series expansion of $P_{n,\mu}(\alpha,\beta;x)$ in $x$ given by Eq.~(\ref{eq:Biconfluent_Heun_expansion}), as well as the second power series expansion
\begin{equation}
P_{m,\nu}(\alpha,\beta;x)=\sum_{l=0}^{m}\frac{A_{l}}{(1+\alpha)_{l}}\frac{x^{l}}{l!}\ .
\label{eq:explicit_form_Biconfluent_Heun_polynomials_y}
\end{equation}
Using Eqs.~(\ref{eq:Biconfluent_Heun_expansion}) and (\ref{eq:explicit_form_Biconfluent_Heun_polynomials_y}), we obtain the following result
\begin{equation}
N_{n}=\left(\frac{\tau}{I}\right)^{1/2}\ ,
\label{eq:N_n_Schrodinger}
\end{equation}
where
\begin{eqnarray}
I & = & \int_{0}^{\infty}x^{2}\mbox{e}^{-x^{2}}P_{n,\mu}(\alpha,\beta;x)P_{m,\nu}(\alpha,\beta;x)dx\nonumber\\
& = & \sum_{k=0}^{n}\sum_{l=0}^{m}\frac{A_{k}A_{l}}{(1+\alpha)_{k}(1+\alpha)_{l}}\frac{1}{k!l!}\int_{0}^{\infty}x^{2+k+l}\mbox{e}^{-x^{2}}dx\ .
\label{eq:I_Schrodinger_1}
\end{eqnarray}
For $n=m$, $\mu=\nu$ and $k=l$, we have
\begin{eqnarray}
I & = & \int_{0}^{\infty}x^{2}\mbox{e}^{-x^{2}}P_{n,\mu}^{2}(\alpha,\beta;x)dx\nonumber\\
& = & \sum_{k=0}^{n}\frac{A_{k}^{2}}{[(1+\alpha)_{k}k!]^{2}}\int_{0}^{\infty}x^{2+2k}\mbox{e}^{-x^{2}}dx\ .
\label{eq:I_Schrodinger_2}
\end{eqnarray}
Since
\begin{equation}
\int_{0}^{\infty}y^{\lambda-1}\mbox{e}^{-\eta y^{u}}dy=\frac{1}{u}\eta^{-\lambda/u}\Gamma\left(\frac{\lambda}{u}\right)\ ,
\label{eq:integral}
\end{equation}
the integral in Eq.~(\ref{eq:I_Schrodinger_2}) is simply
\begin{equation}
I=\sum_{k=0}^{n}\frac{A_{k}^{2}}{[(1+\alpha)_{k}k!]^{2}}\frac{1}{2}\Gamma\left(\frac{3}{2}+k\right)\ .
\label{eq:I_Schrodinger_3}
\end{equation}
From Eqs.~(\ref{eq:N_n_Schrodinger}) and (\ref{eq:I_Schrodinger_3}) we see that apart from an arbitrary complex multiplicative factor of modulus one the normalisation constant $N_{n}$ is given by
\begin{equation}
N_{n}=\left[\sum_{k=0}^{n}\frac{A_{k}^{2}}{[(1+\alpha)_{k}k!]^{2}}\frac{1}{2\tau}\Gamma\left(\frac{3}{2}+k\right)\right]^{-1/2}\ .
\label{eq:normalisation_constant_Schrodinger}
\end{equation}
Using Eq.~(\ref{eq:normalisation_constant_Schrodinger}), we can write the normalized solution of the Schr\"{o}dinger equation in a Newtonian universe as
\begin{eqnarray}
\psi_{n,\mu}(R) & = & \left[\sum_{k=0}^{n}\frac{A_{k}^{2}}{[(1+\alpha)_{k}k!]^{2}}\frac{1}{2\tau}\Gamma\left(\frac{3}{2}+k\right)\right]^{-1/2}\nonumber\\
& \times & \tau R\ \mbox{e}^{-\tau^{2}R^{2}/2}\ \mbox{HeunB}(\alpha,\beta,\alpha+2(n+1),\delta_{\mu}^{n};\tau R)\ .
\label{eq:normalised_Newtonian_Universe_eigenfunctions}
\end{eqnarray}
%
%
\section{Conclusions}
We have presented exact and general solution of the Schr\"{o}dinger equation for a particle (galaxy) moving in a Newtonian universe in the presence of a cosmological constant term. The complete set of solutions, given in terms of the biconfluent Heun functions, satifies the appropriate boundary conditions $\psi(0)=0$ and $\psi(\infty)=0$, and is valid over the range $0 \leq x < \infty$.

In order to get a well-behaved solution, we imposed the polynomial condition on these analytic solutions and thus we obtained the Heun polynomials for the biconfluent case.

An exact expression for the energy spectrum was also obtained. These energy levels are similar to the ones corresponding to a three-dimensional isotropic harmonic oscillator, with an effective frequency given in terms of the cosmological constant.

As a conclusion we emphasize that the wave functions as well as the energy spectrum codifies the presence of a global cosmological force which affects the particles (galaxies) of this Newtonian universe.
%
%
\begin{acknowledgments}
The authors would like to thank Conselho Nacional de Desenvolvimento Cient\'{i}fico e Tecnol\'{o}gico (CNPq) for partial financial support.
\end{acknowledgments}
%
%

%
%

\begin{thebibliography}{99}
\bibitem{Schutz:2009} B. Schutz, \textit{A first course in general relativity}, (Cambridge University Press, New York, 2009).
\bibitem{Bondi:2010} H. Bondi, \textit{Cosmology}, (Dover Publications, INC., New York, 2010).
\bibitem{QJMath.5.64} E. A. Milne, Q. J. Math. \textbf{5}, 64 (1934).
\bibitem{QJMath.5.73} W. H. McCrea and E. A. Milne, Q. J. Math. \textbf{5}, 73 (1934).
\bibitem{AmJPhys.33.105} C. Callan, R. H. Dicke and P. J. E. Peebles, Am. J. Phys. \textbf{33}, 105 (1965).
\bibitem{ProcRSocLondA.206.562} W. H. McCrea, Proc. R. Soc. Lond. A \textbf{206}, 562 (1951).
\bibitem{ProcRSocLondA.149.384} E. T. Whittaker, Proc. R. Soc. Lond. A \textbf{149}, 384 (1935).
\bibitem{PhysRevLett.109.051303} T. Clifton, C. Clarkson and P. Bull, Phys. Rev. Lett. \textbf{109}, 051303 (2012).
\bibitem{AstronAstrophys.571.A15} Planck Collaboration and P. A. R. Ade et al., Astron. Astrophys. \textbf{571}, A15 (2014).
\bibitem{AIPConfProc.743.286} D. Z. Freedman, M. Schnabl and G. W. Gibbons, AIP Conf. Proc. \textbf{743}, 286 (2004).
\bibitem{arXiv:0504072} J. M. Romero and A. Zamora, arXiv:0504072 \textbf{[gr-qc]} (2005).
\bibitem{ProcRSoc.A.463.503} B. Bramson, Proc. R. Soc. A \textbf{463}, 503 (2007).
\bibitem{IntJTheorPhys.47.455} H. T. Elze, Int. J. Theor. Phys. \textbf{47}, 455 (2008).
\bibitem{ISRNMathPhys.2013.509316} C. Kiefer, ISRN Math. Phys. \textbf{2013}, 509316 (2013).
\bibitem{MathAnn.33.161} K. Heun, Math. Ann. \textbf{33}, 161 (1889).
\bibitem{JPhysAMathGen.19.3527} B. Leaute and G. Marcilhacy, J. Phys. A: Math. Gen. \textbf{19}, 3527 (1986).
\bibitem{AnnPhys.347.130} F. Caruso, J. Martins and V. Oguri, Ann. Phys. (NY) \textbf{347}, 130 (2014).
\bibitem{arXiv:1101.0471v5} M. Horta\c csu, in \textit{Mathematical Physics: Proceedings of the 13th Regional Conference}, edited by U. Camci and I. Semiz (World Scientific, Singapore, 2013), pp.~23--39; e-print arXiv:1101.0471v5 \textbf{[math-ph]} (2013).
%
\bibitem{RevBrasEnsFis.36.3310} H. S. Vieira and V. B. Bezerra, Rev. Bras. Ens. Fis. \textbf{36}, 3310 (2014).
\bibitem{Ronveaux:1995} A. Ronveaux, \textit{Heun's differential equations}, (Oxford University Press, New York, 1995).
\bibitem{AnnSocSciBruxelles.92.151} A. Decarreau, P. Maroni and A. Robert, Ann. Soc. Sci. Bruxelles \textbf{92}, 151 (1978).
\bibitem{BullSocRSciLiege.40.13} A. Hautot, Bull. Soc. R. Sci. Li\`{e}ge \textbf{40}, 13 (1969).
\end{thebibliography}
\end{document}